# Photonic topological phase transition on demand

Zhaxylyk A. Kudyshev[*], Alexander V. Kildishev, Alexandra Boltasseva,

and Vladimir M. Shalaev

*School of Electrical and Computer Engineering and Birck Nanotechnology Center,*

*Purdue University, West Lafayette, IN 47907, USA*

**Abstract**: On-demand, switchable phase transitions between topologically non-trivial and trivial photonic states are demonstrated. Specifically, it is shown that integration of a 2D array of coupled ring resonators within a thermal heater array enables unparalleled control over topological protection of photonic modes. Importantly, auxiliary control over spatial phase modulation opens up a way to guide topologically protected edge modes along generated virtual boundaries. The proposed approach can lead to practical realizations of topological phase transitions in many photonic applications, including topologically protected photonic memory/logic devices, robust optical modulators, and switches.

## 1. Introduction

Robust photonic systems for dynamic control of light that are insensitive to fabrication imperfections or disorder are in high demand. While advanced control over the light flow can be achieved in an engineered optical environment, such as photonic crystals [1] or metamaterials [2], these artificial material platforms require precise fabrication techniques, because photonic states in such systems are susceptible to various perturbation channels. One possible way of addressing this problem is to use so-called topologically protected photonic states. This concept has recently been translated from the electronic system to photonics, boosting the interest in a new class of topologically ordered optical systems – so-called photonic topological insulators [3–6]. Such studies provide new insights into the physics of light-matter interaction, and could enable fundamentally new advanced photonic applications [7,8]. The main advantage of using topologically protected states is their robustness to perturbations of photonic states caused by system's imperfections/inhomogeneities. A number of approaches for realizing topologically protected photonic states has been theoretically proposed and experimentally demonstrated [9–13]. Photonic topological insulators have been mainly realized using metamaterials exhibiting a magneto-optical response from engineered meta-atoms, or all-dielectric metamaterials that exploit electric and magnetic resonances of nanoparticles with a high refractive index [14–16]. Variations of photonic crystal structures have been used for the realization of non-trivial topological photonic states as well [17,18]. Another recent approach [19–26] has utilized a system of resonators with a controlled coupling that forms topologically non-trivial frequency gaps with robust edge states. Photonic topological insulators have already shown to be a promising platform for realizing topologically protected lasers [27–29] and a laser cavity of arbitrary shape [30], along with unidirectional waveguides [31], and a promising quantum optics venue [32,33]. The controllable transformation between different topological states in a photonic system is a challenging and important task. Realization of on-demand topological phase transitions will open up unparalleled control over topological protected photonic states, which in turn, can bring in a fundamentally new way of robust optical signal control.

A theoretical possibility of topological transformations between topologically trivial and non-trivial states has been predicted with the tight-binding model framework [34]. In this work, we show that by integrating a system of phase modulators with a standard CMOS-compatible silicon-on-insulator (SOI) technology, it is possible to get unparalleled control over phase difference between clockwise (spin-up) and anticlockwise(spin-down) modes within a unit cell, which defines the level of degeneracy of the modes. The latter fact determines the strength of the synthetic gauge magnetic field control for photons which defines topological uniqueness of the photonic states. We show that this technique enables the realization of topological phase transformation of the photonic system between different topologically protected states. Specifically, we propose to use thermal heating elements for phase modulation in couplers via thermo-optic effect in silicon. Using modulation in couplers allows modifying the accumulated phase of light propagating through the waveguides and hence provides a magnetic field control. We show two examples of the phase transformations: (i) transformation of the system, which leads to re-routing of the protected edge photonic modes around long or short edges of the sample; (ii) spatially distributed topological transformation which enables routing of the topologically protected states along virtual boundaries of complex shapes.

## 2. Phase transitions in photonics topological insulators

- **Tight binding model**

Two-dimensional electron gas in an external magnetic field can be described within nearest neighbors coupling approximation by the Harper-Hofstadter model [35]:

$$H_0 = \sum_{x,y} \hat{a}^\dagger_{x,y}\hat{a}_{x,y} - J\left(\sum_{x,y} \hat{a}^\dagger_{x+1,y}\hat{a}_{x,y}e^{-iy\phi} + \hat{a}^\dagger_{x,y}\hat{a}_{x+1,y}e^{iy\phi} + \hat{a}^\dagger_{x,y+1}\hat{a}_{x,y} + \hat{a}^\dagger_{x,y}\hat{a}_{x,y+1}\right) \quad (1)$$

where $\hat{a}^\dagger_{x,y}$, $\hat{a}_{x,y}$ are the creation (annihilation) operators at a site (x, y). The first term here appears due to the periodic nature of the lattice potential. The second and third terms describe tunneling of the electrons along the $\pm x$-direction, while the last term stands for $\pm y$ direction propagation. The effective tunneling rate is defined as $J$. Due to the presence of the external magnetic field, hopping of electrons in the lattice is accompanied by an additional phase accumulation $\phi$. This Peierls phase is a manifestation of the Aharonov-Bohm phase, proportional to the applied magnetic field [36].

Eigenvalues of the tight-binding Hamiltonian, when plotted against the applied magnetic field, produce a well-known Hofstadter butterfly spectrum (Fig.1a) [37]. For a given co-prime ratio $\alpha_M = \phi/2\pi = p/q$ (p and q prime numbers), there are $q$ allowed bands (filled with non-topologically protected bulk states), and $q-1$ band-gaps. The centermost bands are degenerate at the gauge fields which corresponds to even $q$ values. For a finite size lattice, forbidden band-gaps of the spectrum are populated by topologically protected edge-states, which are robust against backscattering. Such states are confined to the edge of the lattice and immune against geometrical perturbation of the system. Topological order of the system is determined by a topological invariant integer called Chern number $C_m$ ($m$ - integer number). Chern number for square lattice system can be calculated using the Diophantine equations [38]. Figure 1a shows the Hofstadter butterfly spectrum for a 15×15 resonator system. Here, we highlight only band-gaps of interest with the corresponding Chern numbers. However, in reality, there are some band-gaps populated by the edge states that are also determined by the co-prime ratio $p/q$ but not highlighted here [35].

Eigenfunctions of the three different states ($\alpha_M^{(1)} = 0.45, \alpha_M^{(2)} = 0.5, \alpha_M^{(3)} = 0.55$, outlined in Fig. 1a), are shown in Fig.1b-d. Hence, we consider two topologically protected edge states ($\alpha_M^{(1)}$ and $\alpha_M^{(3)}$)

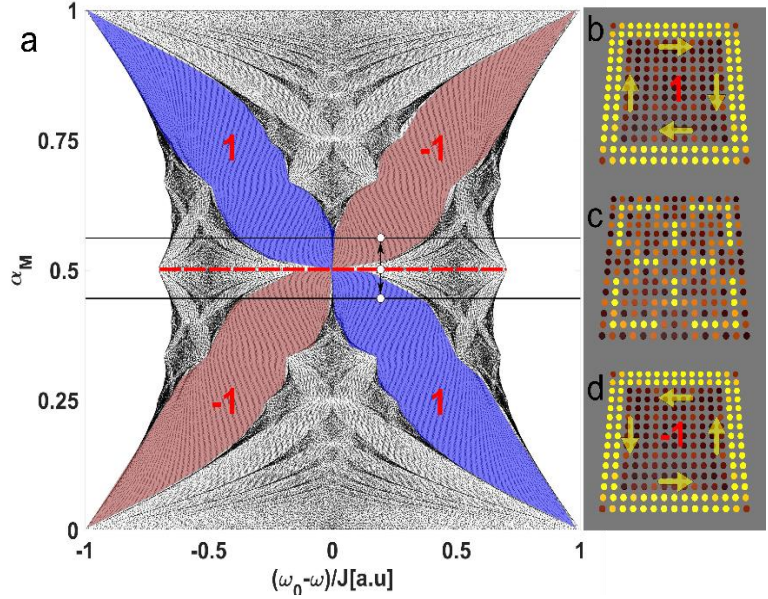

Fig.1 (a) Hofstadter butterfly spectrum for a 15×15 resonator system. Eigenfunctions of the three different states: (b) $\alpha_M^{(1)} = 0.45$ - topologically protected state with unity Chern number $C_{+1}$, (c) $\alpha_M^{(2)} = 0.5$ (trivial state), (d) $\alpha_M^{(3)} = 0.55$ - topologically protected state with negative Chern number $C_{-1}$. Arrows show direction of propagation of probability currents in corresponding edge states.

with opposite unity Chern numbers ($C_{+1}$ and $C_{-1}$) and one trivial state ($\alpha_M^{(2)} = 0.5$). Arrows show the direction of the probability currents propagation in corresponding edge states. We also note that the state corresponding to $\alpha_M^{(2)} = 0.5$, $(p=1, q=2)$ has a trivial topology with two joint bands forming a Dirac cone at the center of the Brillouin zone. This example demonstrates that a reasonable modulation of the external magnetic field leads to a dramatic change in the systems' topology, which in turn allows for achieving unparalleled control over topological protection of the system transport properties. In this study, we extend this concept to photonic systems to achieve control over topologically protected light transport through a topological phase transition concept.

- **Photonic analog of a 2D electron system**

A series of works has previously demonstrated that under certain arrangements, an array of coupled ring resonators enables the realization of a photonic analog of a two-dimensional electron gas in an external magnetic field [21,39,40] (Fig.2a). A unit cell of the array consists of four "site" resonators and four "link" waveguides, forming a rectangular lattice (inset in Fig.2a). The "site" resonators are coupled evanescently to the "link" waveguides, hence providing transfer to their nearest neighbors, while the "link" waveguides are detuned from the resonance wavelength, thus

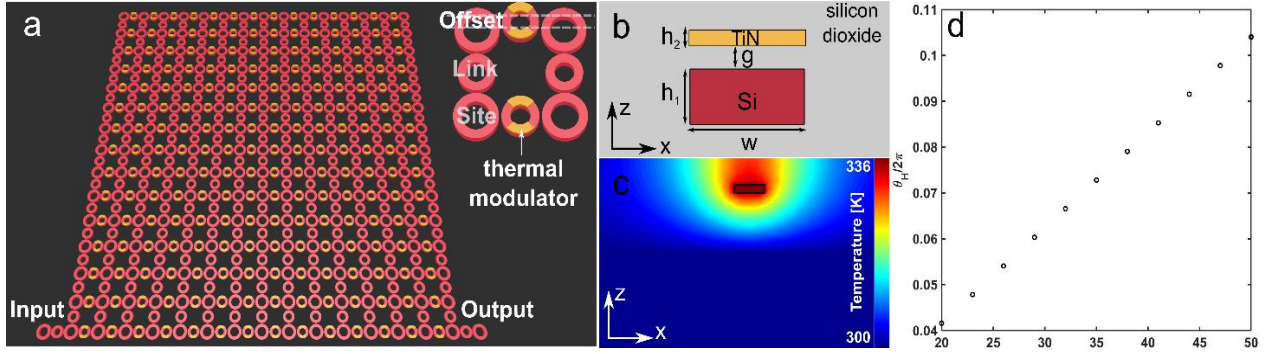

Fig.2 (a) Schematics of the coupled 15×15 coupled ring resonators. Inset shows a unit cell configuration. Resonators made of silicon placed in silicon dioxide environment; (b) Cross section of the link waveguide integrated with thermal heater ($w = 510\,nm, h_1 = 220\,nm, h_2 = 100\,nm, g = 600\,[nm]$); (c) Temperature distribution around waveguide (outlined by black contour) for 20-mW input power; (d) the acquired phase shift (in fractions of $2\pi$) for a heater length of $15\,\mu m$.

making all the energy confined in the site rings. We associate the *clockwise propagating photons* in sites with *spin-up* states and *anticlockwise propagating photons* with *spin-down* states respectively [21]. Degeneracy of these two pseudo-spin states could be achieved by forcing photons to hop in the corresponding directions (forward/backward) to acquire different propagation phase $\phi$. Such phase difference can be achieved by shifting one of the "link" waveguides from the symmetry point by vertical offset $\xi$ (see the inset in Fig.2a). In this case, the accumulated phase is equal to the difference in the optical path for forward and backward propagating photons, $\phi = 2\pi\alpha_M = 4\pi n_{mod}\xi / \lambda$, where $n_{mod}$ is a mode index of the waveguide and $\lambda$ is a free-space wavelength of light. By using the analogy with two-dimensional electron gas, we note that the geometrical offset of the link waveguides in a photonic system causes the same effect as the external magnetic field in a quantum system and can be considered as a source of a synthetic magnetic field for photons. To achieve topological phase transitions in a photonic system, it is necessary to introduce control over propagation phase accumulation within each unit cell of the system. Such control can be achieved by integrating phase modulators into the domain so that the total phase would have the following form, $\phi_{tot} = \phi(\xi) + \theta_H$, where $\phi(\xi)$ is a phase introduced by the geometrical offset, and $\theta_H$ is an additional phase introduced by the modulator.

There are different ways of introducing phase modulation into a silicon-on-insulator waveguide. Thermal effects, as well as nonlinear effects, can be used to achieve the phase modulation of the signal employing external electrical or optical excitation. For instance, optical or electrical nonlinear modulation of transparent conducting oxides (TCOs) can provide an ultrafast response

(up to ~100 fs modulation speed), which can be used for signal modulation. However, TCO-based modulation requires more sophisticated designs for achieving the desired modulation depth as well as reducing unavoidable attenuation. To demonstrate proof-of-concept topological phase transitions in photonic topological insulators with phase modulation of link waveguides, we integrate strongly coupled ring resonators with thermal heaters (see Fig. 2a). A similar approach has been utilized to experimentally measure a topological invariant (winding number) in a 2D photonic system [40]. A cross-section of the link waveguide integrated with a thermal heater is shown in Fig. 2b. All resonators are assumed to have a cross-section of $510 \times 220 \, \text{nm}^2$ that limits their operation regime at the telecommunication wavelength range to single-mode propagation of the transverse electric (TE) mode. We consider the simplest design of the heater, comprising a 100-nm-thick TiN plate placed 600 nm above the modulation region. The TiN heater acts as a uniform heat source, which is used to thermally tune the waveguide enabling the desired phase accumulation $\theta_H$ in the regions of interest. We employ a commercial coupled thermal – optical solver (Lumerical Inc., DEVICE integrated with MODE Solutions) to determine heat induced effective index modulation in the silicon waveguide. Temperature distribution around waveguide induced by an input power of 25 mW is shown in Fig. 2c. In the simulations, the thermo-optic coefficient of the silicon waveguide is taken to be $dn/dt = 1.8 \times 10^{-4}$ K$^{-1}$ [41]. Temperature rise leads to modulation of the modal index of the waveguide, which in turn leads to additional phase accumulation of the transmitted signal. Figure 2d shows phase change as a function of input power, revealing that the acquired phase shift increases linearly with the increasing heater power. We consider the phase accumulation around a working wavelength of 1.3 $\mu m$, under the assumption that the heater covers ~45% of the link waveguide total length (~ 65 $\mu m$). This analysis indicates that the system with the thermal heater provides an additional phase of about $\theta_H \approx 0.2\pi$, which would allow us to realize topological phase transition between three states outlined in Fig. 1. By placing the initial state of the system at $\alpha_M^{(1)}$ via the corresponding geometric offset of link waveguides, input powers of 25 mW and 50 mW applied to each heater would bring the system into $\alpha_M^{(2)}$ (trivial) and $\alpha_M^{(3)}$ (topologically protected) states correspondingly, hence realizing topological phase transformation.

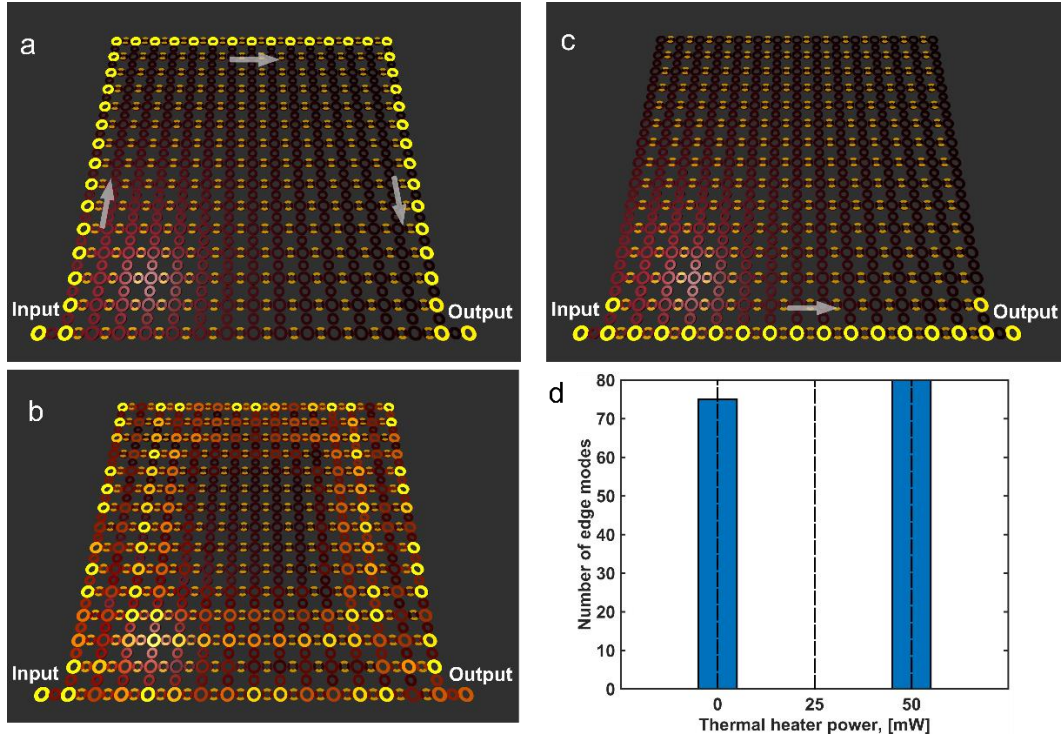

Fig.3 (a)-(c) Field distributions of 15×15 resonator array for three different cases at working wavelength $\lambda = 1.305\,\mu m$ : (a) thermal heater are off, which corresponds to topologically protected state at $\alpha_M^{(1)} = 0.45$ with $C_{+1}$ Chern number; (b) topologically trivial state $\alpha_M^{(2)} = 0.5$ which corresponds to an input power of 25 mW applied to each heater; and (c) topologically protected state $\alpha_M^{(3)} = 0.55$ with , which corresponds to an input power of 50 mW applied to each heater. (d) Number of edge modes corresponding to three different states. Direction of propagation is indicated by arrows.

## 3. Results and discussions

Using the transfer matrix method (outlined in the appendix), we considered a dynamically tunable platform integrating a thermal heater with a photonic system, comprising an array of 15×15 site microring resonators (Fig. 2a). Using a commercial solver (Lumerical MODE Solutions), we first obtained system parameters needed for the transfer matrix method, such as coupling ($\kappa$) and transmission ($\tau$) coefficients with an add/drop filter (a single resonator coupled to two waveguides) as well as two ring resonators coupled through an off-resonant middle ring. This analysis indicates that for achieving a resonant response of the photonic system around 1.3 μm wavelength, the length of the site resonators should be $L_{\mathrm{LR}} = 65\,\mu m$, while an additional length $\eta$ of the link waveguides leads to anti-resonant behavior with the site resonators should be equal to $\eta = 150\,\mathrm{nm}$, making $L_{\mathrm{LR}} = 65.15\,\mu m$. Inter-waveguide spacing of 100 nm together with 120 nm gap between bus input-output waveguides and a coupling length of 12.8 μm yeild coupling coefficients of $\kappa_{ij} = 0.41$ and $\kappa_{\text{in}} = 0.48$. We assume that the thermal heater covers 45% of the tunable link waveguides surface and is placed such that it covers two sides of the link waveguide, which are not coupled to the neighboring site resonators, ensuring that coupling is the same within the photonic lattice (see the inset in Fig. 2a).

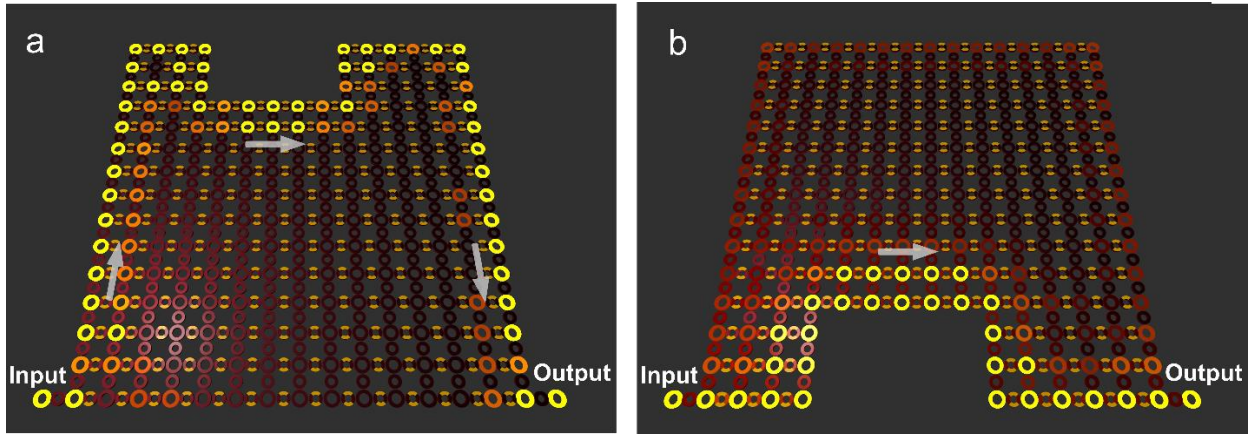

Fig.4 Field distributions inside 15×15 resonator array for topologically protected states with lattice perturbations (defects) at working wavelength $\lambda = 1.305\,\mu m$: (a) topologically protected state $\alpha_M^{(1)}$ with $C_{+1}$ Chern number (thermal heater are off); (b) topologically protected state $\alpha_M^{(3)}$ with $C_{-1}$ Chern number (50-mW applied power).

To enable the gauge magnetic field $\alpha_M^{(1)} = 0.45$ geometric offset of the link waveguides should be $\zeta = 215\,\text{nm}$. This state corresponds to topological protection state $\alpha_M^{(1)}$ with $C_{+1}$ Chern number. Field distribution of 15×15 resonator array at the working wavelength $\lambda = 1.305\,\mu m$ is shown in Fig. 3a. Following the prediction of the tight-binding model, this state enables backscattering-free clockwise propagation of the topologically protected edge mode. To realize topological phase transition which brings the system into the trivial state with $\alpha_M^{(2)} = 0.5$, an input power of 25 mW should be applied to each thermal heater. All of the supported modes at this state are topologically trivial and correspond to bulk modes. The field distribution corresponding to $\alpha_M^{(2)}$-state is depicted in Fig. 3b. By applying an input power of 50 mW, we realize topological phase transition that places the system into the topologically protected ($\alpha_M^{(3)} = 0.55$) state with $C_{-1}$. The topologically protected edge mode at this state is shown in Fig. 3c; it corresponds to counter-clockwise propagating, backscattering-free edge mode. To verify the topological phase transitions, we have counted all the supported edge modes at all three states ( $\alpha_M^{(1)}, \alpha_M^{(2)}$ and $\alpha_M^{(3)}$ ). Figure 3d depicts the statistics of this analysis; as expected $\alpha_M^{(1)}$ and $\alpha_M^{(3)}$ states support a certain number of topologically protected edge states, while there are no edge modes in $\alpha_M^{(2)}$-state. Hence, we show that the photonic topological phase transformation can be enabled through incorporation of additional phase modulation and appropriate optimization of the coupled ring resonator array.

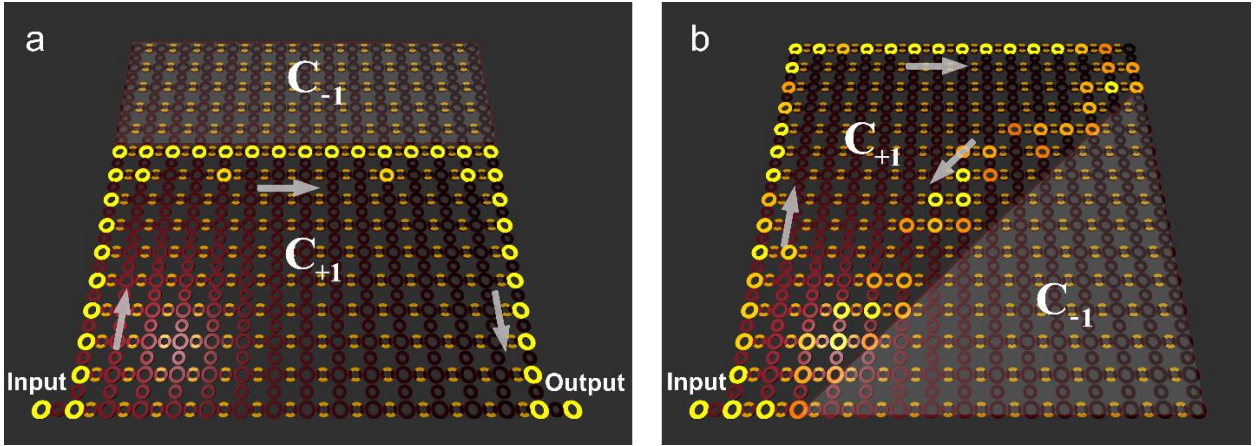

Fig.5 Field distributions inside 15×15 resonator array for spatially distributed applied power (50-mW input power regions outlined by shadowing), which leads to spatially distributed topologically protected states $\alpha_M^{(1)}$ and $\alpha_M^{(3)}$: (a) straight, vertical virtual boundary between topological insulators; (b) an input power of 50 mW applied to each heating element positioned below the main diagonal of the array.

One of the main properties of the topologically protected photonic states is robustness of the edge states against structural disorders and imperfections. To demonstrate this, we consider a photonic lattice with a missing resonator along the edges (Fig. 4). Figure 4a shows the field distribution for the topologically protected edge mode at $\alpha_M^{(1)}$ state (the thermal heaters are off). Figure 4b shows the counter-clockwise propagating topologically protected edge mode at $\alpha_M^{(3)}$ state (an input power of 50 mW). Both cases demonstrate that due to topological protection these modes bypass the impurity without scattering into bulk or backscattering. This is not a case for topologically trivial states suffering from scattering losses, which would dramatically decrease mode transmission.

Spatial control over applied input power to heaters system allows constructing topologically protected virtual interfaces on demand. Distribution of the applied power determines the spatial distribution of the additional phase accumulation $\theta_H$ in the system, which in turn defines the virtual interface between two different photonic topological insulators with $C_{+1}$ and $C_{-1}$ Chern numbers. Figure 5 shows two examples of such spatially distributed topological phase transition. Specifically, Figure 5a shows the case when an input power of 50 mW is applied to the half of the heater domain (6 upper rows of the resonators, outlined by shadowing) forming a virtual flat interface between $C_{+1}$ and

$C_{-1}$ photonic topological insulators. It can be seen, that spatially distributed topological phase transition locks topologically protected edge state into the virtual boundary and force the mode to propagate along a predefined trajectory. Figure 5b shows a more complex routing version when a power of 50 mW is applied to each heating element positioned below the main diagonal of the 15×15 resonator array (outlined by the shadowing). The $C_{+1}$ topology of the $\alpha_M^{(1)}$ state forces the mode to propagate along the long edge of the sample. Once it reaches the virtual boundary, the mode starts to propagate backward following the edge between two insulated states $\alpha_M^{(1)}$ and $\alpha_M^{(3)}$. These two examples demonstrate that spatial control over topology brings a fundamentally new way of on-demand light manipulation.

## 4. Conclusion

In this work, we demonstrated the concept of the tunable topological phase transition realized in a photonic system consisting of a coupled ring resonator array. We showed that by integrating an array of heating elements it is possible to introduce an additional propagation phase, which is accumulated along with the one introduced by a geometrical offset of the link couplers. This approach opens up a new way of precise control over gauge magnetic field, which in turn leads to flexible, on-demand control over the topology of the photonic system. We showed that by defining the spatial distribution of the input power applied to the heaters, we can adjust topological properties, as well as determine the distribution of the virtual boundaries for the topologically protected edge modes. The realization of such control over topology is a crucial step towards topologically protected photonic memory/logic devices, which are in high demand for quantum communication systems and quantum computing. Future studies can be focused on developing topologically protected memory concepts based on the so-called Lieb lattice design, which consists of two dispersive bands that touch a flat middle band (stop-band) via a cone-like dispersion point.

**Appendix: Transfer matrix method for coupled waveguide system**

We have used the transfer matrix formalism, which can be applied to strongly coupled microring resonator systems of interest [42]. The response of the microring resonator array can be obtained by dividing each waveguide into four sections denoted as $\mathbf{a}, \mathbf{b}, \mathbf{c}, \mathbf{d}$ ($\mathbf{m} = [m_1, ..., m_N]^T$, $\mathbf{m} \in (\mathbf{a}, \mathbf{b}, \mathbf{c}, \mathbf{d})$). For the lattice, the number of site resonators is $N_{SR} = N_x \times N_y$, while the number of link resonators is

$N_{LR} = N_x(N_y - 1) + N_y(N_x - 1)$. So, the total length of the vector $\mathbf{m}$ is $N = N_{LR} + N_{SR}$. Coupling between two adjacent resonators is denoted by the field coupling coefficient $\kappa_{ij}$.

The sections are connected to each other by a set of coupling junctions which can be described by the four transfer matrices $\mathbf{M}_j, j \in 1...4$. These N×N matrices are symmetric with the property that if there is coupling between waveguides $i$ and $j$ at the junction $k$ with coupling coefficient $\kappa_{ij}$ then

$$\begin{aligned} \mathbf{M}_k(i,i) &= \mathbf{M}_k(j,j) = \tau_{ij}\Phi A \\ \mathbf{M}_k(i,j) &= \mathbf{M}_k(j,i) = i\kappa_{ij}\Phi A \end{aligned} \quad (2)$$

here $\tau_{ij}^2 + \kappa_{ij}^2 = 1$. If waveguide $i$ is uncoupled at the junction k then $\mathbf{M}_k(i,i) = 1$. $\Phi, A$ are the phase and attenuation coefficient, respectively. Both are determined by the type of a microring resonators coupled through the junction: a) for "site" resonators $\Phi = \exp(-i\beta L_{SR}/4)$ and $A = \exp(-\alpha L_{SR}/4)$; b) for "link" waveguides $\Phi = \exp(-i\beta L_{LR}/4)$ and $A = \exp(-\alpha L_{LR}/4)$; and c) for tunable "link" coupler $\Phi = \exp(-i\beta(L_{LR}/4 \pm \zeta))\exp(-i\theta_H)$ and $A = \exp(-i\alpha L_{LR}/4)$. Here $L_{SR}$ and $L_{LR}$ are lengths of the "site" resonators and "link" waveguides, $\beta$ and $\alpha$ are the propagation constant and field decay coefficient of silicon waveguide. Tunable "link" waveguides have additional phase accumulation due to vertical displacement $\zeta$ from center position, the sign of which is determined by propagation direction of the mode inside the junction. Spin-up states correspond to plus phase accumulation, while spin-down states acquire negative phase due to displacement.

By introducing the coupling matrix $\mathbf{M} = \mathbf{M}_4\mathbf{M}_3\mathbf{M}_2\mathbf{M}_1$, evolution equation of the field in the microring resonator array can be written as:

$$\mathbf{a} = \mathbf{LMa} + \mathbf{s} \quad (3)$$

here $\mathbf{L} = \text{diag}[\tau_i, 1, ..., 1, \tau_o]$ is a diagonal matrix representing the output/input bus-to-ring couplings, and $\mathbf{s} = [i\kappa_i, 0, ..., 0]^T$ is an input field array. Here we assume that resonator 1 and N are coupled to input and output bus waveguides via input and output coupling coefficients. Using Eq. (3) the closed form solution for field vector can be written as:

$$\mathbf{a} = \frac{\mathbf{s}}{\mathbf{I} - \mathbf{LM}} \qquad (4)$$

where $\mathbf{I}$ is the identity matrix. Once we determine electromagnetic field distribution within one section, we can determine all remaining section field vectors using coupling matrixes $\mathbf{M}_j, j \in 1..4$.

**ACKNOWLEDGMENTS**

This work was supported in part from U.S. Department of Energy (DE-SC0017717), AFOSR (FA9550-18-1-0002), DARPA Extreme Optics and Imaging (EXTREME) program (HR00111720032) and ONR (N00014-18-1-2481).

**References:**

[1]   J. D. Joannopoulos, P. R. Villeneuve, and S. Fan, Nature **386**, 143 (1997).
[2]   V. M. Shalaev, Nat. Photonics **1**, 41 (2007).
[3]   F. D. M. Haldane and S. Raghu, Phys. Rev. Lett. **100**, (2008).
[4]   S. Raghu and F. D. M. Haldane, Phys. Rev. A - At. Mol. Opt. Phys. **78**, (2008).
[5]   S. Weimann, M. Kremer, Y. Plotnik, Y. Lumer, S. Nolte, K. G. Makris, M. Segev, M. C. Rechtsman, and A. Szameit, Nat. Mater. (2016).
[6]   F. Gao, Z. Gao, X. Shi, Z. Yang, X. Lin, H. Xu, J. D. Joannopoulos, M. Soljaaia, H. Chen, L. Lu, Y. Chong, and B. Zhang, Nat. Commun. **7**, (2016).
[7]   L. Lu, J. D. Joannopoulos, M. Soljacic, M. Soljačić, and M. Soljacic, Nat. Photonics **8**, 821 (2014).
[8]   A. B. Khanikaev and G. Shvets, Nat. Photonics **11**, 763 (2017).
[9]   J. Noh, S. Huang, K. P. Chen, and M. C. Rechtsman, Phys. Rev. Lett. **120**, (2018).
[10]  L. Lu, J. D. Joannopoulos, and M. Soljačić, Nat. Phys. **12**, 626 (2016).
[11]  M. I. Shalaev, S. Desnavi, W. Walasik, and N. M. Litchinitser, New J. Phys. **20**, (2018).

[12]  X. Cheng, C. Jouvaud, X. Ni, S. H. Mousavi, A. Z. Genack, and A. B. Khanikaev, Nat. Mater. **15**, 542 (2016).

[13]  H. Zhou, C. Peng, Y. Yoon, C. W. Hsu, K. A. Nelson, L. Fu, J. D. Joannopoulos, S. Marin, and B. Zhen, Science (80-. ). **359**, 1009 (2018).

[14]  A. B. Khanikaev, S. Hossein Mousavi, W. K. Tse, M. Kargarian, A. H. MacDonald, and G. Shvets, Nat. Mater. **12**, 233 (2013).

[15]  T. Ma, A. B. Khanikaev, S. H. Mousavi, and G. Shvets, Phys. Rev. Lett. **114**, (2015).

[16]  W. Gao, M. Lawrence, B. Yang, F. Liu, F. Fang, B. Beri, J. Li, and S. Zhang, Phys. Rev. Lett. **114**, (2015).

[17]  S. Barik, H. Miyake, W. Degottardi, E. Waks, and M. Hafezi, New J. Phys. **18**, (2016).

[18]  L.-H. Wu and X. Hu, Phys. Rev. Lett. **114**, 223901 (2015).

[19]  K. Fang, Z. Yu, and S. Fan, Nat. Photonics **6**, 782 (2012).

[20]  M. Hafezi, M. D. Lukin, and J. M. Taylor, New J. Phys. **15**, (2013).

[21]  M. Hafezi, S. Mittal, J. Fan, A. Migdall, and J. M. Taylor, Nat. Photonics **7**, 1001 (2013).

[22]  A. Blanco-Redondo, I. Andonegui, M. J. Collins, G. Harari, Y. Lumer, M. C. Rechtsman, B. J. Eggleton, and M. Segev, Phys. Rev. Lett. **116**, (2016).

[23]  Y. Plotnik, M. A. Bandres, S. Stützer, Y. Lumer, M. C. Rechtsman, A. Szameit, and M. Segev, Phys. Rev. B **94**, (2016).

[24]  M. A. Bandres, M. C. Rechtsman, and M. Segev, Phys. Rev. X **6**, (2016).

[25]  M. C. Rechtsman, Y. Plotnik, J. M. Zeuner, D. Song, Z. Chen, A. Szameit, and M. Segev, Phys. Rev. Lett. **111**, (2013).

[26]  M. C. Rechtsman, J. M. Zeuner, Y. Plotnik, Y. Lumer, D. Podolsky, F. Dreisow, S. Nolte, M. Segev, and A. Szameit, Nature **496**, 196 (2013).

[27]  M. A. Bandres, S. Wittek, G. Harari, M. Parto, J. Ren, M. Segev, D. N. Christodoulides, and M. Khajavikhan, Science (80-. ). **359**, (2018).

[28]  G. Harari, M. A. Bandres, Y. Lumer, M. C. Rechtsman, Y. D. Chong, M. Khajavikhan, D. N. Christodoulides, and M. Segev, Science (80-. ). **359**, (2018).

[29]  H. Zhao, P. Miao, M. H. Teimourpour, S. Malzard, R. El-Ganainy, H. Schomerus, and L. Feng, Nat. Commun. **9**, 981 (2018).

[30]  B. Bahari, A. Ndao, F. Vallini, A. El Amili, Y. Fainman, and B. Kanté, Science (80-. ). 1 (2017).

[31] Z. Wang, Y. Chong, J. D. Joannopoulos, and M. Soljačić, Nature **461**, 772 (2009).

[32] J. Perczel, J. Borregaard, D. E. Chang, H. Pichler, S. F. Yelin, P. Zoller, and M. D. Lukin, Phys. Rev. Lett. **119**, (2017).

[33] S. Barik, A. Karasahin, C. Flower, T. Cai, H. Miyake, W. DeGottardi, M. Hafezi, and E. Waks, Science (80-. ). **359**, 666 (2018).

[34] D. Leykam, S. Mittal, M. Hafezi, and Y. D. Chong, Phys. Rev. Lett. **121**, (2018).

[35] D. R. Hofstadter, Phys. Rev. B **14**, 2239 (1976).

[36] Y. Aharonov and D. Bohm, Phys. Rev. **115**, 485 (1959).

[37] F. T. P. Harper, The Hofstadter Model and Other Fractional Chern Insulators, 2015.

[38] M. Kohmoto, Phys. Rev. B **39**, 11943 (1989).

[39] S. Mittal, J. Fan, S. Faez, A. Migdall, J. M. Taylor, and M. Hafezi, Phys. Rev. Lett. **113**, (2014).

[40] S. Mittal, S. Ganeshan, J. Fan, A. Vaezi, and M. Hafezi, Nat. Photonics **10**, 180 (2016).

[41] H. W. Icenogle, B. C. Platt, and W. L. Wolfe, Appl. Opt. **15**, 2348 (1976).

[42] A. Tsay and V. Van, IEEE J. Quantum Electron. **47**, 997 (2011).